\documentclass[11pt]{svjour2}
\usepackage{amsmath}
\usepackage{amsfonts}
\usepackage{epsfig}
\usepackage{color}

\newcommand{\bfe}{\mbox{\boldmath $e$}}
\newcommand{\eee}{\vec{e}}
\newcommand{\JJ}{\tilde{J}}
\newcommand{\JJJ}{\vec{J}}
\newcommand{\KKK}{\vec{\vec{K}}}
\newcommand{\bfp}{\mbox{\boldmath $p$}}
\newcommand{\QQ}{\tilde{Q}}
\newcommand{\QQQ}{\vec{Q}}
\newcommand{\vvv}{\vec{v}}
\newcommand{\bfv}{\mbox{\boldmath $v$}}

\newcommand{\ppphi}{\vec{\phi}}

\begin{document}

\title{Being, Becoming and the Undivided Universe:
A Dialogue between Relational Blockworld and the Implicate Order
Concerning the Unification of Relativity and Quantum Theory}
\institute{Michael Silberstein \at Department of Philosophy \\ Elizabethtown College \\
Elizabethtown, PA  17022 \\ Tel. 717-361-1253\\
\email{silbermd@etown.edu} \and W.M. Stuckey \at Department of
Physics
\\ Elizabethtown College \\ Elizabethtown, PA  17022 \\ Tel.
717-361-1436 \\
\email{stuckeym@etown.edu}
\and Timothy McDevitt \at Department of Mathematical Sciences \\
Elizabethtown College \\ Elizabethtown, PA 17022 \\ Tel.
717-361-1337\\
\email{mcdevittt@etown.edu}}\maketitle

\begin{abstract}
In this paper two different approaches to unification will be
compared, Relational Blockworld (RBW) and Hiley's implicate order.
Both approaches are monistic in that they attempt to derive matter
and spacetime geometry `at once' in an interdependent and background
independent fashion from something underneath both quantum theory
and relativity. Hiley's monism resides in the implicate order via
Clifford algebras and is based on process as fundamental while RBW's
monism resides in spacetimematter via path integrals over graphs
whereby space, time and matter are co-constructed per a global
constraint equation. RBW's monism therefore resides in being
(relational blockworld) while that of Hiley's resides in becoming
(elementary processes). Regarding the derivation of quantum theory
and relativity, the promises and pitfalls of both approaches will be
elaborated. Finally, special attention will be paid as to how
Hiley's process account might avoid the blockworld implications of
relativity and the frozen time problem of canonical quantum gravity.
\end{abstract}

\thispagestyle{empty}

\section{Introduction}
\begin{quotation}Listening not to me but to the Logos it is wise to agree that
all things are one.
        -Heraclitus\end{quotation}

\begin{quotation}There remains, then, but one word by which to express the
[true] road: Is. And on this road there are many signs that What Is
has no beginning and never will be destroyed: it is whole, still,
and without end. It neither was nor will be, it simply is-now,
altogether, one, continuous ...
        -Parmenides\end{quotation}

\label{section1}
\subsection{Modeling Fundamental Reality and Ultimate Explanation: A Schism in Physics}
There has been a very long standing debate in Western philosophy and
physics regarding the following three pairs of choices about how
best to model the universe: 1) the fundamentality of being versus
becoming, 2) monism versus atomism and 3) algebra versus geometry
broadly construed; more generally, which of the myriad formalisms
will be most unifying.

Regarding 1, from very early on Western thinkers have generally
assumed that everything can be explained. Perhaps the cosmological
argument for the existence of God is the classic example of such
thinking. In that argument Leibniz appeals to a version of the
principle of sufficient reason (PSR) which states\cite{melamed} ``no
fact can be real or existing and no statement true without a
sufficient reason for its being so and not otherwise.'' Leibniz uses
the principle to argue that the sufficient reason for the ``series
of things comprehended in the universe of creatures must exist
outside this series of contingencies and is found in a necessary
being that we call God''\cite{melamed}. While physics dispensed with
appeals to God at some point, it did not jettison PSR, merely
replacing God with fundamental dynamical laws, e.g., as anticipated
for a Theory of Everything (TOE), and initial conditions (the big
bang or some condition leading to it). In keeping with everyday
experience a very early assumption of Western physics--reaching its
apotheosis with Newtonian mechanics--is that the fundamental
phenomena in need of explanation are \emph{motion} and \emph{change
in time}, so explanation will involve dynamical laws most
essentially.

In the quest to unify all of physics, it is the combination of PSR
plus the dynamical perspective writ large (call it dynamism) that
has in great part motivated the particular kind of unification being
sought, i.e., the search for a TOE, quantum gravity (QG) and the
like. Therefore, almost all attempts to unify relativity and quantum
theory opt for becoming (dynamism) as fundamental in some form or
another. Such theories may deviate from the norm by employing
radical new fundamental dynamical entities (branes, loops, ordered
sets, etc.), but the game is always dynamical, broadly construed
(vibrating branes, geometrodynamics, sequential growth process,
etc).

However, it is also important to note that from fairly early on in
Western physics there have also been adynamical explanations that
focused on the role of the future in explaining the past as well as
the reverse, such as integral (as opposed to differential) calculus
and various least action principles of the sort Richard Feynman
generalized to produce the path integral approach to quantum
mechanics. And of course there are the various adynamical
constraints in physics such as conservation laws and the symmetries
underlying them that constrain if not determine the various
equations of motion. But nonetheless, dynamism is still the reigning
assumption in physics.

Dynamism then encompasses three claims: A) the world, just as
appearances and the experience of time suggest, evolves or changes
in time in some objective fashion, B) the best explanation for A
will be be some dynamical law that ``governs'' the evolution of the
system in question, and C) the fundamental entities in a TOE will
themselves be dynamical entities evolving in some space however
abstract, e.g., Hilbert space. In spite of the presumption of
dynamism, those who want fundamental explanation in physics to be
dynamical and those who want a world that evolves in time in some
objective fashion, face well-known problems concerning: 1) the
possible blockworld implications of relativity (both special and
general) and 2) canonical QG, the quantization of a generally
covariant classical theory leading to ``frozen time.'' As for
whether relativity (both special and general) implies a blockworld,
there is much debate\cite{savitt}. Regarding special relativity
(SR), many of us have argued\cite{peterson} that given certain
widely held innocuous assumptions and the Minkoswski formulation,
special relativity does indeed imply a blockworld. In the words of
Geroch\cite{geroch}:
\begin{quotation}There is no dynamics within space-time itself:
nothing ever moves therein; nothing happens; nothing changes.
In particular, one does not think of particles as moving through
space-time, or as following along their world-lines. Rather, particles
are just in space-time, once and for all, and the world-line represents,
all at once, the complete life history of the particle.\end{quotation}

In addition there is the problem of time in canonical general
relativity (GR). That is, in a particular Hamiltonian formulation of
GR the reparametrization of spacetime is a gauge symmetry.
Therefore, all genuinely physical magnitudes are constants of
motion, i.e., they don't change over time. In short, change is
merely a redundancy of the representation.

Finally, the problem of frozen time in canonical QG (unification of
GR and quantum field theory) is that if the canonical variables of
the theory to be quantized transform as scalars under time
reparametrizations, which is true in practice because they have a
simple geometrical meaning, then ``the Hamiltonian is (weakly) zero
for a generally covariant system''\cite{henneaux}. The result upon
canonical quantization is the famous Wheeler-DeWitt equation, void
of time evolution. While it is too strong to say a generally
covariant theory must have H = 0, there is no well-developed theory
of quantum gravity that has avoided it to date\cite{kiefer1}. It is
supremely ironic that the dynamism and unificationism historically
driving physics led us directly to blockworld and frozen time.

Two basic reactions to this tension between blockworld and frozen
time on the one hand and dynamism on the other are to either embrace
the former and show that at least the appearances of dynamism, if
not the substance, can be maintained with resources intrinsic to
relativity or the particular QG scheme in
question\cite{savitt}\cite{kiefer2}, or reject the former whether
conceptually or formally and attempt to construct a fundamental
theory that has something definitively dynamical at bottom. The idea
is to somehow make time or change fundamental in some way, as
opposed to merely emergent as in the case of string theory or an
illusion as in the case of Wheeler-Dewitt. Smolin, for example,
suggests a radically ``neo-Heraclitean'' solution wherein change and
becoming are fundamental in that axiomatic dynamical laws, the
values of constants that figure in those laws and configuration
space itself evolve in time or meta-time\cite{smolin}. Though he
does not necessarily frame it this way, Smolin is advocating for
something like a fundamentally Whiteheadian process conception of
reality, a process-based physics where change or flux itself is
fundamental. In doing so, Smolin joins Bohm and Hiley who have been
advocating such an approach for many decades\cite{bohm1}.

However, what isn't clear is if Smolin appreciates what a radical
departure a process-conception of reality is from atomism wherein
reality has some fundamental dynamical building blocks (atoms,
particles, waves, strings, loops, etc.) from which everything else
is constructed, determined or realized. This brings us to choice
point number 2, atomism versus monism. Despite all the tension that
quantum theory has created for atomism as originally conceived, most
physicists still assume there is something fundamentally entity-like
at bottom, however strange it may be by classical lights. But on the
process view, potentia, activity, flux or change itself is
fundamental, not entities/things changing in time such as particles
or strings. In this monistic physics (what Bohm and Hiley call
``undivided wholeness''), all talk of such dynamical entities would
emerge from, and be derived from, the more fundamental flux together
with, and inseparably from, spacetime in a background independent
fashion (the formal question remains of course as to how this move
would resolve for example the problem of frozen time). Thus Bohm and
Hiley are constructing a monistic model wherein ``the whole is prior
to its parts, and thus views the cosmos as fundamental, with
metaphysical explanation dangling downward from the
One''\cite{schaffer}. However, the motivations for a process-based
physics are not exclusively physical, but are also driven by the
desire to have fundamental concepts of physical time correspond with
time and change as experienced such that time as experienced isn't
merely a subjective psychological feature of humans with no clear
physical correspondence. Following Price\cite{price}, the key
elements to time as experienced are: \emph{objectively dynamical}
(flow or flux-like), \emph{present moment} objectively
distinguished, and \emph{objective direction}.

This brings us to choice point number 3, algebra versus geometry
broadly construed. There is a dizzying array of formalisms at work
in physics. In quantum mechanics alone we have matrix mechanics,
Schr\"{o}dinger dynamics, Clifford algebras, and path integrals, to
name a few, and in quantum field theory (QFT) we have canonical
quantization, covariant quantization, path integral method,
Becchi-Rouet-Stora-Tyupin (BRST) approach, Batalin-Vilkovisky (BV)
quantization, and Stochastic quantization\cite{kaku1}. When we get
to QG and unification the list is even longer and more
diverse\cite{kiefer3}. Throughout history there have always been
differences of opinion, some pragmatic and some principled, about
which formalism(s) best models fundamental physical reality. Indeed,
one of the striking things about the state of unification is the
heterogeneity of formal approaches and the lack of consensus despite
the juggernaut of string theory and its progeny. Hiley for example,
likes to say that in his program, geometry (spacetime) is derived
from algebra (process), rather than the other way
around\cite{hiley1}. Other approaches, such as ours, proceed along
something closer to the opposite direction. Hiley enumerates several
advantages to using orthogonal Clifford algebras in quantum
mechanics: 1) they provide a mathematical hierarchy of nested
algebras in which to naturally embed the Dirac, Pauli and
Schr\"{o}dinger particles, 2) the approach is fully algebraic, which
allows a more general approach to quantum phenomena, 3) because it
is an algebraic theory, it provides a natural mathematical setting
for the Heisenberg `matrix' mechanics, 4) because it is
representation free, it avoids the use of multiple indices on
spinors , and 5) it removes the \emph{ad hoc} features of the
earlier attempts to extend the Bohm approach to spin and
relativity\cite{hiley2}. But, what is interesting from the
perspective of foundations of physics is that while there is no
necessary connection between a formalism and a particular model or
metaphysical interpretation, we see that theorists sometimes pick a
formalism based in part on their prior metaphysical biases and
background beliefs about the nature of reality, in addition to other
physical and formal considerations pertaining to unification such as
those Hiley gives above. For example, one of the main reasons Hiley
adopts an algebraic approach at bottom is that he thinks algebra can
better model process whereas the geometrization of time in
relativity leads exactly to blockworld, a conception of reality he
rejects as too static. Indeed, at least on the surface it is hard to
imagine a cosmology less comforting to a process conception of
reality than blockworld or H = 0. At any rate, what should now be
clear is that each of our three choice points has implications for
the others.

\subsection{Prelude: RBW versus the Implicate Order}

In this paper two different approaches to unification will be
compared, the Relational Blockworld (RBW) emphasizes being over
becoming formally and conceptually, while the Implicate Order of
Hiley emphasizes the converse. RBW has something closer to geometry
at bottom (discrete graphical structure) while Hiley has Clifford
algebras as fundamental. Each of these programs was originally
spawned by two diametrically opposed solutions to foundational
issues in non-relativistic quantum mechanics (NRQM) and QFT, rather
than starting life as models of
QG\cite{stuckey1}\cite{silberstein}\cite{bohm2}. As we will see,
while both are cast in the monistic spirit, Hiley's monism resides
in Bohm's implicate order and is based on process while RBW's monism
resides in ``spacetimematter,'' whereby space, time and matter are
co-constructed per a global constraint equation; RBW's monism
therefore resides in being while that of Hiley resides in becoming.
Both these programs have proposed new formalisms for quantum physics
and are in the process of extrapolating their approaches to
unification and quantum
gravity\cite{hiley1}\cite{stuckey2}\cite{hiley3}\cite{hiley4}\cite{hiley5}.

The Implicate Order of Hiley extends Bohmian mechanics to the
relativistic regime and unites spacetime geometry and material
processes, as he doesn't want things happening in a background
spacetime but wants to ``start from something more primitive from
which both geometry and material process unfold
together''\cite{hiley6}. That which he considers ``more primitive''
is elementary process. Hiley calls the fundamental process/potentia
the ``holomovement'' and it has two intertwined aspects, the
``implicate order'' (characterized algebraically) and all the
physics derived from it, such as spacetime geometry, the ``explicate
(or manifest) order.'' The holomovement is thus the whole ground
form of existence, which contains orders that are both implicate and
explicate, wherein the latter expresses aspects of the former. Hiley
reduces the Clifford algebra $C_{4,1}$ to $C_{1,3}$ whence he
derives the vector space of M4 by mapping the Dirac gamma matrices
to the orthonormal vectors spanning $V_{1,3}$ of M4. He then defines
Bohm momentum and energy densities in the Dirac equation in analogy
with his earlier work with Bohm\cite{hiley7}. From the perspective
of the implicate order, rather than point particles being evolved in
time aided by instantaneous updating by the quantum potential or
pilot wave, the fundamental evolution is one of processes that give
rise to explicate structures (``moments'' or ``durons'') extended in
space and time. In short, particles and pilot waves are not
fundamental but are at best emergent from the implicate order (see
section \ref{section4}). The irony is not lost on Hiley that the
Bohm and Hiley work on interpreting NRQM has done more than perhaps
any other interpretation to bolster a particle ontology and a
``mechanical'' conception of the quantum modeled on an analogy with
classical mechanics\cite{goldstein}. Indeed, as we will see in
section \ref{section2}, much of Hiley's later work is trying to get
out from under such a pseudo-classical model and emphasize the
undivided wholeness instead.

However, in order for Hiley to finish his program, presumably, he
will need to accommodate any Lagrangian, not just that of the Dirac
equation. For example, he will need to compute cross sections for
the various collision experiments of high energy physics. If he
proceeds along the lines of ``current algebra''\cite{kaku2}, as
suggested by his approach to
date\cite{hiley1}\cite{hiley4}\cite{hiley5}, perhaps he could
produce a Bohmian explanation for why the commutators between some
currents in the Standard Model do not close, producing the so-called
Schwinger terms. But, even if he were able to find an algebra of
process for the Standard Model that provided Bohm momenta and
energies for all the particles, he would still have only ``a first
approximation to the true theory of subatomic
particles''\cite{kaku3}, since the Standard Model is plagued with
twenty-some-odd free parameters. He would be in the same boat as
everyone else, needing to account for the free parameters of the
Standard Model and include gravity (see section \ref{section3}). The
point is that Hiley would have to join the ranks of theorists who
are still looking for a `super-algebra' whence the Lagrangian
unifying the Standard Model and gravity.

\begin{figure}
\includegraphics[height=30mm]{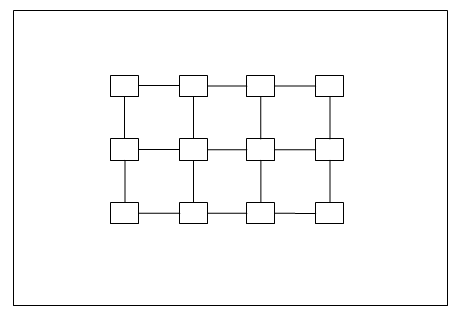} \hspace{10mm}
\includegraphics[height=60mm]{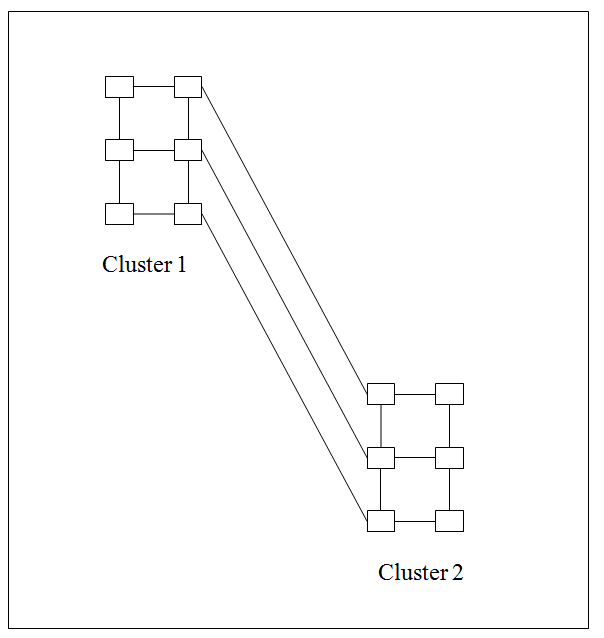}
\caption{(a) Topological Graph - This spacetimematter graph depicts
four sources, i.e., the columns of squares. The graph's actional
$\KKK/2+\JJJ$, such that $\KKK\cdot \vvv \propto \JJJ$,
characterizes the graphical topology, which underwrites a partition
function $Z$ for spatiotemporal geometries over the graph. (b)
Geometric Graph - The topological graph of (a) is endowed with a
particular distribution of spatiotemporal geometric relations, i.e.,
link lengths as determined by the field values $Q$. Clusters 1 \& 2 are the result of this
geometric process for a particular distribution of field values
$Q$.} \label{fig1}
\end{figure}

\begin{figure}
\includegraphics[height=50mm]{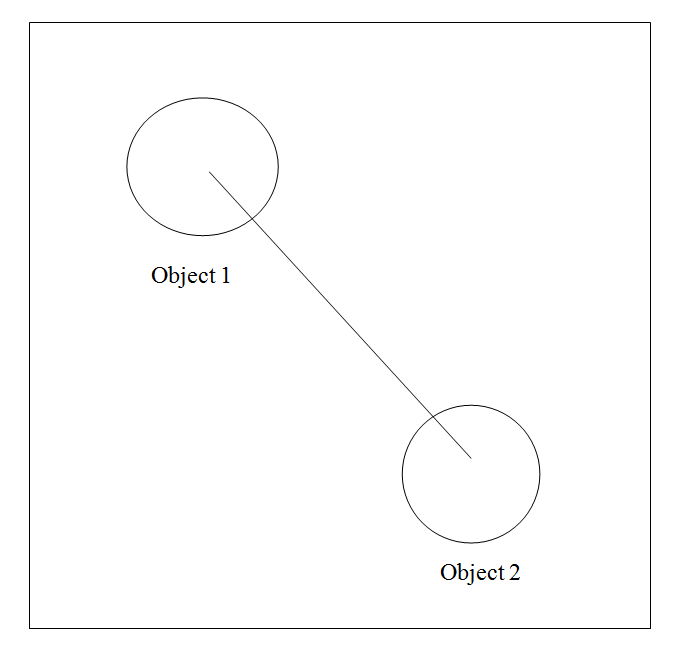} \hspace{10mm}
\includegraphics[height=50mm]{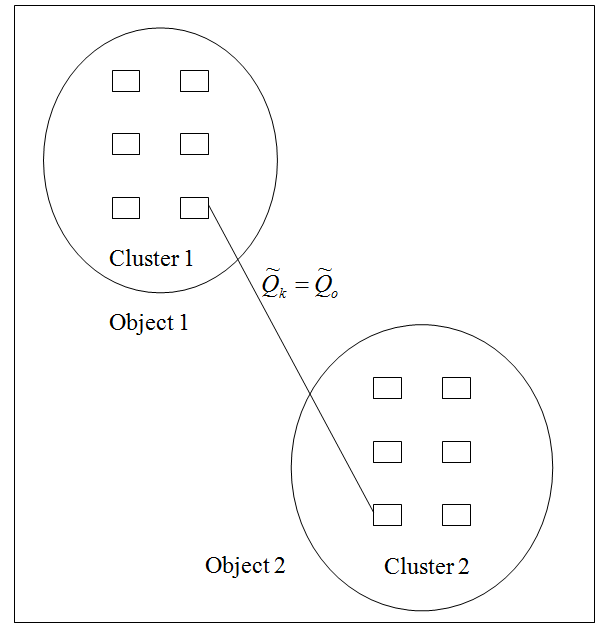}
\caption{(a) Classical Physics - Classical Objects result when the
most probable field values $\QQQ_0$ yield spatiotemporally localized
Clusters 1 \& 2 as in Figure \ref{fig1}b. The lone link in this
figure represents the average of the link lengths obtained via the
most probable field values $\QQQ_0$. The most probable values
$\QQQ_0$ are found via $\KKK \cdot \QQQ_0 = \JJJ$, so this is the
origin of classical physics. (b) Quantum Physics - A particular
outcome $\QQ_0$ of a quantum physics experiment allows one to
compute the $k^{\mbox{th}}$ link length of the geometric graph in
the context of the classical Objects comprising the experiment,
e.g., Source, beam splitters, mirrors, and detectors. The partition
function provides the probability of this particular outcome, i.e.,
$\displaystyle P(\QQ_k = \QQ_0) = \frac {Z(\QQ_k = \QQ_0)}{Z}$.}
\label{fig2}
\end{figure}

As with Hiley's implicate order, our account of quantum physics,
which we call the Relational Blockworld (RBW), is based on a form of
monism, i.e., the unity of space, time and matter at the most
fundamental level. We call this fundamental unity
``spacetimematter'' and use it to recover dynamical or process-like
classical physics only statistically. Thus, we do not attempt to
derive geometry from algebra but in a sense, the other way round
(see section \ref{section2}). In order to appreciate how GR
``emerges'' on our view, it is important to understand that, unlike
Hiley's account, our approach is fundamentally adynamical and
acausal, again, in contrast also to other fundamental theories
attempting to quantize gravity (M-theory, loop quantum gravity,
causets, etc.). According to RBW, as we will explain in detail in
section \ref{section2}, quantum physics is the continuous
approximation of a more fundamental, discrete graph theory whereby
the transition amplitude Z is not viewed as a sum over all paths in
configuration space, but is a measure of the symmetry of the
difference matrix and source vector of the discrete graphical action
for a 4D process (Figure \ref{fig1}a). We have proposed that the
source vector and difference matrix of the discrete action in the
path integral be constructed from boundary operators on the graph so
as to satisfy an adynamical constraint equation we call the
``self-consistency criterion'' (SCC), (see section \ref{section2}
for details). While itself adynamical, the SCC guarantees the graph
will produce divergence-free classical dynamics in the appropriate
statistical limit (Figure \ref{fig2}a), and provides an acausal
global constraint that results in a self-consistent co-construction
of space, time and matter that is \emph{de facto} background
independent. Thus, in RBW one has an acausal, adynamical unity of
``spacetimematter'' at the fundamental level that results
statistically in the causal, dynamical ``spacetime + matter'' of
classical physics. This graphical amalgam of spacetimematter is the
basis for all quantum phenomena as viewed in a classical context
(Figure \ref{fig2}b), that is, we represent this unity of
spacetimematter with 4D graphs constructed per the SCC, and a
Wick-rotated Z provides a partition function for the distribution of
graphical relations responsible statistically for a particular
classical process (Figures \ref{fig1} and \ref{fig2}).

Thus, RBW provides a wave-function-epistemic account of quantum
mechanics with a time-symmetric explanation of interference via
acausal global constraints\cite{silberstein}. Quantum physics is
simply providing a distribution function for graphical relations
responsible for the experimental equipment and process from
initiation to termination. So, while according to some such as
Bohmian mechanics, EPR-correlations and the like evidence
superluminal information exchange (quantum non-locality), and
according to others such correlations represent non-separable
quantum states (quantum non-separability), per RBW these phenomena
are actually evidence of the deeper graphical unity of
spacetimematter responsible for the experimental set up and process,
to include outcomes\cite{stuckey1}\cite{silberstein}. RBW is
therefore integral calculus thinking writ
large\cite{stuckey1}\cite{stuckey2}.

As regards the ``emergence'' or derivation of GR from RBW (see
section \ref{section3}), since we recover classical physics in terms
of the ``average spacetime geometry'' over the graphical unity of
spacetimematter, our discrete average/classical result is a modified
Regge calculus \footnote{Interestingly, in direct correspondence,
Hiley noted that he and Bohm had considered Regge calculus, but
found it emphasized the `structure'  too much and lost the notion of
`process'. By turning to the notion of an `algebra', Hiley found he
could keep the structure aspect, but emphasize more the process.}.
Ordinary Regge calculus is a discrete approximation to GR where the
discrete counterpart to Einstein's equations is obtained from the
least action principal on a 4D graph\cite{misner1}. This generates a
rule for constructing a discrete approximation to the spacetime
manifold of GR using small, contiguous 4D graphical `tetrahedra'
called ``simplices.'' The smaller the legs of the simplices, the
better one may approximate a differentiable manifold via contiguous
simplices. Our proposed modification of Regge calculus (and,
therefore, GR) requires all simplex legs contain non-zero
stress-energy contributions (per spacetimematter), so our simplices
can be both large and non-contiguous. Consequently, per RBW, GR is
seen as a continuous approximation to a modified Regge calculus
wherein the simplices can be large and non-contiguous.

Clearly, Hiley's Implicate Order and RBW differ formally (algebraic
vs path integral) and conceptually (process-oriented vs adynamical).
The monistic character of Hiley's process-oriented approach is
housed in the implicate order, i.e., the Clifford algebra. That
which we observe (the explicate order) is a projection from the
implicate order. Thus, the implicate order accounts for EPR
correlations, which appear to require quantum non-locality (as in
Bohmian mechanics) and/or non-separability in the explicate order of
spacetime. The monistic character of RBW is housed in
spacetimematter which underwrites the spacetime + matter classical
world of our observations. Thus spacetimematter accounts for EPR
correlations, which appear to require quantum non-locality and/or
non-separability in the spacetime + matter of our classical
perspective\cite{stuckey1}\cite{silberstein}. Therefore, both
approaches want to explain such observed quantum phenomena from a
more fundamental theory underneath quantum theory itself, though
these are quite opposing fundamental theories. More specifically,
both approaches want to derive GR and quantum theory from something
more fundamental in a background independent fashion such that the
explanation for quantum entanglement and EPR correlations, rather
than creating tensions with spacetime and relativity, requires
neither non-locality nor non-separability in spacetime. Rather, such
quantum effects (their phenomenology) are explained at the more
fundamental level whether graphical or algebraic. In section
\ref{section2} we provide a brief overview of Hiley's implicate
order (details are already published elsewhere) and a technical
overview of RBW. In sections \ref{section3} and \ref{section4} we
explore their respective prospects for providing progress in the
quest for unification and quantum gravity, and discuss their
perspectives on dynamism.

\section{Quantum Field Theory: Implicate Order Versus RBW}
\label{section2}
\subsection{Hiley's Implicate Order}

Hiley has issued the following challenge\cite{hiley6}:
\begin{quotation}Since the advent of general relativity in which matter and
geometry codetermine each other, there is a growing realisation that
starting from an \emph{a priori} given manifold in which we allow
material processes to unfold is, at best, limited. Can we start from
something more primitive from which both geometry and material
process unfold together? The challenge is to find a formalism that
would allow this to happen.\end{quotation} Hiley then refers to
Bohm's early attempt\cite{hiley6}:
\begin{quotation}David Bohm introduced the notion of a discrete structural
process in which he takes as basic, not matter or fields in
space-time, but a notion of `structure process' from which the
geometry of space-time and its relationship to matter emerge
together providing a way that could underpin general relativity and
quantum theory.\end{quotation}

While Hiley's view may seem radical to some, he is not alone in
appreciating what quantum theory and GR have wrought and what their
unification may require\cite{rovelli}:\begin{quotation}General
relativity (GR) altered the classical understanding of the concepts
of space and time in a way which...is far from being fully
understood yet. QM challenged the classical account of matter and
causality, to a degree which is still the subject of controversies.
After the discovery of GR we are no longer sure of what is spacetime
and after the discovery of QM we are no longer sure of what matter
is. \emph{The very distinction between space-time and matter is
likely to be ill-founded}....I think it is fair to say that today we
do not have a consistent picture of the physical world. [italics
added]\end{quotation}

With regard to QFT, Hiley's own response to his challenge employs
``Clifford algebras taken over the reals'' to provide ``a coherent
mathematical setting for the Bohm formalism.'' In particular, he is
concerned with finding the Bohm momentum and energy in a
relativistic theory, i.e., the Dirac theory, since a common
criticism of Bohm's view is that it cannot be applied in the
relativistic domain. Early attempts by Bohm at making his approach
relativistically invariant focused on the conserved Dirac current
$J^{\mu}=\langle\bar{\Psi}|\gamma^{\mu}|\Psi\rangle$ which results
from global gauge invariance $\psi\rightarrow e^{i\theta}\psi$.
Hiley finds another conserved current associated with the Dirac
particle, the energy-momentum density current
$2iT^{\mu0}=\psi^\dag(\partial^{\mu}\psi)-(\partial^{\mu}\psi^\dag)\psi$
which results from invariance under spacetime translations. Hiley
argues that this energy-momentum density current is the relativistic
counterpart to Bohm energy and momentum for the Schr\"{o}dinger
particle, $E_B=-\partial_tS$ and $\vec{p}_B=\nabla S$. This differs
from the standard treatment of the Dirac particle whereby the
energy-momentum current is only integrated for global conservation
of energy and momentum. In standard field theory, the Dirac current
is stressed, since it couples to the gauge field. Hiley's view leads
to a curious split of the Dirac particle into a `Bohm' part and a
`gauge' part. The split is unique to the relativistic regime, as
there is no such split for the Schr\"{o}dinger or Pauli particles.
So, what does this relativistic dual nature suggest?

Hiley speculates it is indicative of a composite or extended nature
of the Dirac particle. While this idea would apply to baryons, as
they are understood as extended and composed of quarks, it would not
appear relevant to leptons, which are understood as point-like and
fundamental. And what, for example, would we expect for a Bohmian
explanation of the twin-slit experiment using Dirac particles? Would
the resulting interference pattern be explained by trajectories for
the energy-momentum density current in analogy with the Bohmian
Schr\"{o}dinger particle? If so, how would the change in this
interference pattern in the Aharanov-Bohm experiment be explained?
Since it is the Dirac current that couples to the gauge field and it
is the gauge field that is responsible for the Aharanov-Bohm shift
in the interference pattern, we would expect the Bohmian
trajectories to adhere in some respect to the Dirac current. We
suspect that this is indicative of an underlying problem, i.e.,
trying to understand relativistic quantum phenomena in the context
of a particular Lorentz frame, as is done by generating his minimal
left ideal with the idempotent $\epsilon_1=(1+\gamma^0)/2$. We don't
see any problem with his suggested correspondence between his Dirac
energy-momentum density current and its non-relativistic, non-spin
limit of the Bohm energy and momentum for the Schr\"{o}dinger
particle, i.e., $\rho E_B = T^{00}$ and $\rho P_B = T^{k0}$.
However, the fact that it is the energy-momentum density current
that makes this correspondence, rather than the Dirac current,
suggests to us a breakdown in the Bohmian view (quantum potential
defined per a particular Lorentz frame), as would be expected when
going to the relativistic regime.

Regardless of whether or not the notion of Bohmian trajectories can
be preserved in the relativistic regime, Hiley's implicate order
does offer a process-based approach to quantum physics via ``a
hierarchy of Clifford algebras which fit naturally the physical
sequence: Twistors $\rightarrow$ relativistic particle with spin
$\rightarrow$ non-relativistic particle with spin $\rightarrow$
non-relativistic particle without spin''\cite{hiley8}. And this
approach does unite spacetime geometry and material process via the
primitive notion of process algebra. What is unique about the shadow
manifolds (explicate order) that are projected from his Clifford
algebras is that they lead to an equivalence class of Lorentz
observers, rather than a single Minkowski spacetime manifold (M4).
Any particular Lorentz frame serves as the base space for a Clifford
bundle. Assuming this base space is a flat Riemannian manifold M,
Hiley constructs a derivative D from space-like derivatives on M and
the generators of his Clifford bundle. Thus defined, D is a
connection on M and the momentum operator of quantum mechanics
(Schr\"{o}dinger, Pauli, Dirac equations). He then uses this D to
construct a Hamiltonian whence ``the two dynamical equations that
form the basis of the Bohm approach to quantum mechanics - a
Louville type conservation of probability equation and a quantum
Hamilton-Jacobi equation''\cite{hiley9}. While it may seem like a
weakness that he produces shadow manifolds rather than M4, we see
this as a potential advantage in dealing with the problems of
blockworld and ``frozen time,'' as we will discuss in section
\ref{section4}. For now, we simply point out the obvious challenge,
i.e., he must find a connection with curvature for the tangent space
bundle to the base space manifold so as to recover GR. He speculates
this might be done by analyzing phase information in the exchange of
light signals, since ``the Moyal algebra for relating phase
information can be obtained from a deformed Poisson algebra, which
is obtained via the hidden Heisenberg algebra''\cite{hiley10}. As he
has not begun this project, we can offer only limited speculation on
such an attempt in section \ref{section3}.

Our more general concern is about Hiley's motivation for wanting to
obtain a complete relativistic version of the Bohm model for the
Dirac particle, given that he clearly rejects the fundamentality of
particles and pilot (guide) waves, they are emergent at best.
Consider the following passages from Hiley:

\begin{quotation}We strive to find the elementary objects, the quarks,
the strings, the loops and the M-branes from which we try to
reconstruct the world. Surely we are starting from the wrong
premise. Parker-Rhodes (1981) must be right, so too is Lou Kauffman
(1982)! We should start with the whole and then make distinctions.
Within these distinctions we can make finer distinctions and so
on\cite{hiley11}.\end{quotation}

\begin{quotation}In this paper we want to draw specific attention to a sixth
advantage, namely, that it allows us to apply Clifford algebras to
the Bohm approach outlined in Bohm and Hiley. In fact it provides,
for the first time, an elegant, unified approach to the Bohm model
of the Schr\"{o}dinger, Pauli, and Dirac particles, in which we no
longer have to appeal to any analogy to classical mechanics to
motivate the approach as was done by Bohm in his original
paper\cite{hiley12}.\end{quotation}

When Hiley speaks of analogies to classical mechanics, not only is
he jettisoning point particles as fundamental but also the wave
function and apparently the guide wave:

\begin{quotation}In our
approach, the information normally encoded in the wave function is
already contained within the algebra itself, namely, in the elements
of its minimal left ideals\cite{hiley13}.\end{quotation}

\begin{quotation}Thus we see that at no stage is it necessary to
appeal to classical mechanics and therefore there is no need to
identify the classical action with the phase to motivate the
so-called `guidance' equation $\vec{p}=\nabla S$ as was done in
Bohm's original work\cite{callaghan}.\end{quotation}

\begin{quotation}Then it is not diffcult to show that this again reduces, in the non-relativistic
limit, to the Bohm momentum found in the Pauli case and reduces
further, if the spin is suppressed, to the well-known
Schr\"{o}dinger expression $P_B = \nabla S$. This condition is
sometimes known as the guidance condition, but here we have no
`waves', only process, so this phrase is inappropriate in this
context\cite{hiley14}.\end{quotation}

\begin{quotation} Thus by choosing $\alpha = \frac{1}{2}$ we see that our $\rho P_j$ is simply the momentum
density. Furthermore it also means that $\vec{P} = \vec{p_B}$, the
Bohm momentum. Because this can be written in the form $\vec{p_B} =
\nabla S$. Some authors call this the `guidance' condition, but here
it is simply a bilinear invariant and any notion of `guidance' is
meaningless\cite{hiley15}.\end{quotation}

It seems to us that there has always been a tension in Bohm and
Hiley's ``undivided wholeness'' and the pseudo-classical Bohmian
mechanics conceived as a modal interpretation of NRQM with particles
communicating instantaneously with one another, especially in a
relativistic setting. Why spend so much energy trying to recover a
relativistic Bohmian version of the Dirac particle complete with
particle trajectories when such particles and the guidance wave are
at best emergent, and the wave function is merely epistemic?  In the
earlier work it was thought that the Dirac current would provide a
means of calculating particle trajectories\cite{hiley15a}. In Hiley
and Callaghan's recent work they show that the Dirac current is in
fact different from the Bohm energy-momentum current, leaving them
with two different sets of trajectories\cite{hiley16}; again, all of
which raising the question whether Bohmian trajectories can be
recovered in the relativistic case after all. But even if such
trajectories can be recovered, what's the point of trying to
establish that the Bohmian model is relativistically invariant when
Hiley rejects the fundamentality of, if not realism about, that very
model? If it's the monism \emph{a la} process that matters most to
Hiley, then recovering ordinary quantum mechanics or QFT from the
algebraic base is sufficient, nothing is added by recovering a
relativistically Bohmian mechanics as the latter is just a competing
interpretation of quantum mechanics, one that only makes sense to
pursue if you take seriously point particles and pilot waves, which
apparently Hiley does not. Furthermore, it isn't enough to render
Bohmian mechanics Lorentz invariant, it must also be explained how
the non-locality in that model can be squared with the relativity of
simultaneity. Presumably this problem would get solved by Hiley at
the level of the implicate order as a kind of conspiracy theory, but
again, then why bother with recovering Bohmian trajectories and the
like? In the next section we will see that these problems don't
arise for RBW because that model makes a much cleaner break from the
ontology of particles and wave functions even at the level of
ordinary quantum mechanics in spacetime.

\subsection{RBW and Spacetimematter}

We believe the real issue is the fact that QFT involves the
quantization of a classical field\cite{wallace1} when one would
rather expect QFT to originate independently of classical field
theory, the former typically understood as fundamental to the
latter. Herein we propose a new, fundamental origin for QFT.
Specifically, we follow the possibility articulated by
Wallace\cite{wallace2} that, ``QFTs as a whole are to be regarded
only as approximate descriptions of some as-yet-unknown deeper
theory,'' which he calls ``theory X,'' and we propose a new discrete
path integral formalism over graphs for ``theory X'' underlying QFT.
Accordingly, sources $\JJJ$ , space and time are self-consistently
co-constructed per a graphical self-consistency criterion (SCC)
based on the boundary of a boundary principle\cite{misner2} on the
graph ($\partial_1\cdot
\partial_2 = 0$)\footnote{In a graphical representation of QFT, part of $\JJJ$
represents field disturbances emanating from a source location
(Source) and the other part represents field disturbances incident
on a source location (sink).}. We call this amalgam
``spacetimematter.'' The SCC constrains the difference matrix and
source vector in Z, which then provides the probability for finding
a particular source-to-source relationship in a quantum experiment,
i.e., experiments which probe individual source-to-source relations
(modeled by individual graphical links) as evidenced by discrete
outcomes, such as detector clicks. Since, in QFT, all elements of an
experiment, e.g., beam splitters, mirrors, and detectors, are
represented by interacting sources, we confine ourselves to the
discussion of such controlled circumstances where the empirical
results evidence individual graphical links\footnote{Hereafter, all
reference to ``experiments'' will be to ``quantum experiments.''}.
In this approach, the SCC ensures the source vector is
divergence-free and resides in the row space of the difference
matrix, so the difference matrix will necessarily have a nontrivial
eigenvector with eigenvalue zero, a formal characterization of gauge
invariance. Thus, our proposed approach to theory X provides an
underlying origin for QFT, accounts naturally for gauge invariance,
i.e., via a graphical self-consistency criterion, and excludes
factors of infinity associated with gauge groups of infinite volume,
since the transition amplitude $Z$ is restricted to the row space of
the difference matrix and source vector.

While the formalism we propose for theory X is only suggestive, the
computations are daunting, as will be evident when we present the
rather involved graphical analysis underlying the Gaussian
two-source amplitude which, by contrast, is a trivial problem in its
QFT continuum approximation. However, this approach is not intended
to replace or augment QFT computations. Rather, our proposed theory
X is fundamental to QFT and constitutes a new program for physics,
much as quantum physics relates to classical physics. Therefore, the
motivation for our theory X is, at this point, conceptual and while
there are many conceptual arguments to be made for our
approach\cite{stuckey1}\cite{silberstein}, we restrict ourselves
here to the origins of gauge invariance and QFT.

\subsubsection{The Discrete Path Integral Formalism} We understand
the reader may not be familiar with the path integral formalism, as
Healey puts it\cite{healey}, ``While many contemporary physics texts
present the path-integral quantization of gauge field theories, and
the mathematics of this technique have been intensively studied, I
know of no sustained critical discussions of its conceptual
foundations.'' Therefore, we begin with an overview and
interpretation of the path integral formalism, showing explicitly
how we intend to use ``its conceptual foundations.'' We employ the
discrete path integral formalism because it embodies a 4Dism that
allows us to model spacetimematter. For example, the path integral
approach is based on the fact that\cite{feynman} ``the [S]ource will
emit and the detector receive,'' i.e., the path integral formalism
deals with Sources and sinks as a unity while invoking a description
of the experimental process from initiation to termination. By
assuming the discrete path integral is fundamental to the
(conventional) continuum path integral, we have a graphical basis
for the co-construction of time, space and quantum sources via a
self-consistency criterion (SCC). We will then show how the
graphical amalgam of spacetimematter underlies QFT.

\subsubsection{Path Integral in Quantum Physics}

 In the conventional path integral formalism as used by Zee\cite{zee1} for non-relativistic quantum mechanics (NRQM)
 one starts with the amplitude for the propagation from the initial point in configuration space $q_I$ to
 the final point in configuration space $q_F$ in time $T$ via the unitary operator $e^{-iHT}$,
 i.e., $\displaystyle \left \langle q_F \left | e^{-iHT} \right | q_I \right \rangle$. Breaking the
 time $T$ into $N$ pieces $\delta t$ and inserting the identity between each pair of operators $e^{-iH\delta t}$ via
 the complete set $\int dq | q \rangle \langle q | =1$  we have
 \begin{multline*}
 \left \langle q_F \left | e^{-iHT} \right | q_I \right \rangle = \left [ \prod_{j=1}^{N-1} \int dq_j \right ]
 \left \langle q_F \left | e^{-iH\delta t} \right | q_{N-1} \right \rangle
 \left \langle q_{N-1} \left | e^{-iH\delta t} \right | q_{N-2} \right \rangle
 \ldots \\
 \left \langle q_2 \left | e^{-iH\delta t} \right | q_1 \right \rangle
 \left \langle q_1 \left | e^{-iH\delta t} \right | q_I \right \rangle.
 \end{multline*}
 \begin{equation*}
 \left \langle q_2 \left | e^{-iH\delta t} \right | q_1 \right \rangle
 \left \langle q_1 \left | e^{-iH\delta t} \right | q_I \right \rangle.
 \end{equation*}
With $H=\hat{p}^2/2m + V(\hat{q})$ and $\delta t \rightarrow 0$ one
can then show that the amplitude is given by
\begin{equation}
\left \langle q_F \left | e^{-iHT} \right | q_I \right \rangle =
\int Dq(t) \exp \left [ i \int_0^T dt L(\dot{q},q) \right ],
\label{eqn1}
\end{equation}
where $L(\dot{q},q) = m \dot{q}^2/2-V(q)$ . If $q$ is the spatial
coordinate on a detector transverse to the line joining Source and
detector, then $\displaystyle \prod_{j=1}^{N-1}$ can be thought of
as $N-1$ ``intermediate'' detector surfaces interposed between the
Source and the final (real) detector, and $\int dq_j$ can be thought
of all possible detection sites on the $j^{\mbox{th}}$ intermediate
detector surface. In the continuum limit, these become $\int Dq(t)$
which is therefore viewed as a ``sum over all possible paths'' from
the Source to a particular point on the (real) detector, thus the
term ``path integral formalism'' for conventional NRQM is often
understood as a sum over ``all paths through space.''

To obtain the path integral approach to QFT one associates $q$ with
the oscillator displacement at a {\em particular point} in space
($V(q) = kq^2/2$). In QFT, one takes the limit $\delta x \rightarrow
0$ so that space is filled with oscillators and the resulting
spatial continuity is accounted for mathematically via $q_i(t)
\rightarrow q(t,x)$, which is denoted $\phi(t,x)$ and called a
``field.'' The QFT transition amplitude $Z$ then looks like
\begin{equation}
Z = \int D\phi \exp \left [ i \int d^4 x L( \dot{\phi}, \phi ) \right ]
\label{eqn2}
\end{equation}
where $L(\dot{\phi},\phi) = (d\phi)^2/2 - V(\phi)$ . Impulses $J$
are located in the field to account for particle creation and
annihilation; these $J$ are called ``sources'' in QFT and we have
$L(\dot{\phi},\phi) = (d\phi)^2/2 - V(\phi) + J(t,x) \phi(t,x)$,
which can be rewritten as $L(\dot{\phi},\phi) = \phi D \phi/2 +
J(t,x) \phi(t,x)$, where $D$ is a differential operator. In its
discrete form (typically, but not necessarily, a hypercubic
spacetime lattice), $D \rightarrow \KKK$ (a difference matrix),
$J(t,x)\rightarrow \JJJ$ (each component of which is associated with
a point on the spacetime lattice) and $\phi \rightarrow \QQQ$ (each
component of which is associated with a point on the spacetime
lattice). Again, part of $\JJJ$ represents field disturbances
emanating from a source location (Source) and the other part
represents field disturbances incident on a source location (sink)
in the conventional view of path integral QFT and, in particle
physics, these field disturbances are the particles. We will keep
the partition of $\JJJ$ into Sources and sinks in our theory X, but
there will be no vacuum lattice structure between the discrete set
of sources. The discrete counterpart to (\ref{eqn2}) is
then\cite{zee2}
\begin{equation}
Z = \int \ldots \int dQ_1 \ldots dQ_N \exp \left[ \frac {i}{2} \QQQ
\cdot \KKK \cdot\QQQ + i \JJJ \cdot\QQQ \right ]. \label{eqn3}
\end{equation}
In conventional quantum physics, NRQM is understood as $(0+1)-$dimensional QFT.

\subsubsection{Our Interpretation of the Path Integral in Quantum
Physics}

 We agree that NRQM is to be understood as $(0+1)-$dimensional QFT, but point out
 this is at conceptual odds with our derivation of (\ref{eqn1}) when $\int Dq(t)$
 represented a sum over all paths in space, i.e., when $q$ was understood as a
 location in space (specifically, a location along a detector surface). If NRQM
 is $(0+1)-$dimensional QFT, then $q$ is a field displacement at a single location
 in space. In that case, $\int Dq(t)$ must represent a sum over all field values at a particular
 point on the detector, not a sum over all paths through space from the Source to a particular
 point on the detector (sink). So, how {\em do} we relate a point on the detector (sink) to the Source?

In answering this question, we now explain a formal difference
between conventional path integral NRQM and our proposed approach:
our links only connect and construct discrete sources $\JJJ$, there
are no source-to-spacetime links (there is no vacuum lattice
structure, only spacetimematter). Instead of $\delta x \rightarrow
0$, as in QFT, we assume $\delta x$ is measureable for (such) NRQM
phenomenon. More specifically, we propose starting with (\ref{eqn3})
whence (roughly) NRQM obtains in the limit $\delta t \rightarrow 0$,
as in deriving (\ref{eqn1}), and QFT obtains in the additional limit
$\delta x \rightarrow 0$, as in deriving (\ref{eqn2}). The QFT limit
is well understood as it is the basis for lattice gauge theory and
regularization techniques, so one might argue that we are simply
{\em clarifying} the NRQM limit where the path integral formalism is
not widely employed. However, again, we are proposing a discrete
starting point for theory X, as in (\ref{eqn3}). Of course, that
discrete spacetime is fundamental while ``the usual continuum theory
is very likely only an approximation''\cite{Feinberg} is not new.

\subsubsection{Discrete Path Integral is Fundamental}

The version of theory X we propose is a discrete path integral over
graphs, so  (\ref{eqn3}) {is not a discrete approximation of
(\ref{eqn1}) \& (\ref{eqn2})}, but rather {\em (\ref{eqn1}) \&
(\ref{eqn2}) are continuous approximations of  (\ref{eqn3})}. In the
arena of quantum gravity it is not unusual to find discrete
theories\cite{loll} that are in some way underneath spacetime theory
and theories of ``matter'' such as QFT, e.g., causal dynamical
triangulations\cite{ambjorn}, quantum graphity\cite{konopka} and
causets\cite{sorkin1}. While these approaches are interesting and
promising, the approach taken here for theory X will look more like
Regge calculus quantum gravity (see Bahr \& Dittrich \cite{bahr} and
references therein for recent work along these lines) modified to
contain no vacuum lattice structure.

Placing a discrete path integral at bottom introduces conceptual and
analytical deviations from the conventional, continuum path integral
approach. Conceptually, (\ref{eqn1}) of NRQM represents a sum
over all field values at a particular point on the detector, while
(\ref{eqn3}) of theory X is a mathematical machine that measures
the ``symmetry'' (strength of stationary points) contained in the
core of the discrete action
\begin{equation}
\frac 12 \KKK + \JJJ
\label{eqn4}
\end{equation}
This core or {\em actional} yields the discrete action after
operating on a particular vector $\QQQ$ (field). The actional
represents a {\em fundamental/topological, 4D description of the
experiment} and $Z$ is a measure of its symmetry\footnote{In its
Euclidean form, which is the form we will use, $Z$ is a partition
function.}. For this reason we prefer to call $Z$ the symmetry
amplitude of the 4D experimental configuration. Analytically,
because we are {\em starting} with a discrete formalism, we are in
position to mathematically explicate trans-temporal identity,
whereas this process is unarticulated elsewhere in physics. As we
will now see, this leads to our proposed self-consistency criterion
(SCC) underlying $Z$.

\subsubsection{Self-Consistency Criterion} Our use of a
self-consistency criterion is not without precedent, as we already
have an ideal example in Einstein's equations of GR. Momentum, force
and energy all depend on spatiotemporal measurements (tacit or
explicit), so the stress-energy tensor cannot be constructed without
tacit or explicit knowledge of the spacetime metric (technically,
the stress-energy tensor can be written as the functional derivative
of the matter-energy Lagrangian with respect to the metric). But, if
one wants a ``dynamic spacetime'' in the parlance of GR, the
spacetime metric must depend on the matter-energy distribution in
spacetime. GR solves this dilemma by demanding the stress-energy
tensor be ``consistent'' with the spacetime metric per Einstein's
equations. For example, concerning the stress-energy tensor, Hamber
and Williams write\cite{hamber}, ``In general its covariant
divergence is not zero, but consistency of the Einstein field
equations demands $\nabla^{\alpha} T_{\alpha \beta} = 0$ .'' This
self-consistency hinges on divergence-free sources, which finds a
mathematical underpinning in $\partial
\partial  = 0$. So, Einstein's equations of GR are a mathematical
articulation of the boundary of a boundary principle at the
classical level, i.e., they constitute a self-consistency criterion
at the classical level, as are quantum and classical
electromagnetism\cite{misner3}\cite{wise}. We will provide an
explanation for this fact later, but essentially the graphical SCC
of our theory X gives rise to continuum counterparts in QFT and
classical field theory.

In order to illustrate the discrete mathematical co-constuction of
space, time and sources $\JJJ$, we will use graph theory \emph{a la}
Wise\cite{wise} and find that $\partial_1\cdot \partial_1^T$, where
$\partial_1$ is a boundary operator in the spacetime chain complex
of our graph satisfying $\partial_1\cdot \partial_2 = 0$ , has
precisely the same form as the difference matrix in the discrete
action for coupled harmonic oscillators. Therefore, we are led to
speculate that $\KKK \propto\partial_1\cdot \partial_1^T$. Defining
the source vector  $\JJJ$ relationally via $\JJJ \propto
\partial_1\cdot \eee$ then gives tautologically per $\partial_1\cdot
\partial_2 = 0$ both a divergence-free $\JJJ$ and $\KKK\cdot \vvv
\propto \JJJ$, where $\eee$ is the vector of links and $\vvv$ is the
vector of vertices. $\KKK\cdot \vvv \propto \JJJ$ is our SCC
following from $\partial_1\cdot\partial_2 = 0$, and it defines what
is meant by a self-consistent co-construction of space, time and
divergence-free sources $\JJJ$, thereby constraining $\KKK$ and
$\JJJ$ in $Z$. Thus, our SCC provides a basis for the discrete
action and supports our view that (\ref{eqn3}) is fundamental to
(\ref{eqn1}) \& (\ref{eqn2}), rather than the converse.
Conceptually, that is the basis of our discrete, graphical path
integral approach to theory X. We now provide the details.

\subsubsection{The General Approach}

Again, in theory X, the symmetry amplitude $Z$ contains a discrete
action constructed per a self-consistency criterion (SCC) for space,
time and divergence-free sources $\JJJ$. As introduced above and
argued later below, we will codify the SCC using $\KKK$ and $\JJJ$;
these elements are germane to the transition amplitude $Z$ in the
Central Identity of Quantum Field Theory\cite{zee3},
\begin{equation}
Z = \int D \ppphi \exp \left [ - \frac 12 \ppphi \cdot \KKK \cdot \ppphi - V(\ppphi) + \JJJ \cdot \ppphi \right ] \\
= \exp \left [ -V \left ( \frac {\delta}{\delta J} \right ) \right ] \exp \left [\frac 12 \JJJ \cdot \KKK^{-1} \cdot \JJJ \right ].
\label{eqn5}
\end{equation}
While the field is a mere integration variable used to produce $Z$,
it must reappear at the level of classical field theory. To see how
the field makes it appearance per theory X, consider (\ref{eqn5}) for the simple Gaussian theory ($V(\phi) = 0$). On a
graph with $N$ vertices, (\ref{eqn5}) is
\begin{equation}
Z = \int_{-\infty}^{\infty} \ldots \int_{-\infty}^{\infty} dQ_1
\ldots dQ_N  \exp \left [-\frac 12 \QQQ \cdot \KKK \cdot \QQQ + \JJJ
\cdot \QQQ \right ] \label{eqn6}
\end{equation}
with a solution of
\begin{equation}
Z = \left ( \frac {(2\pi)^N}{\det \KKK} \right )^{1/2} \exp \left [\frac 12 \JJJ \cdot \KKK^{-1} \cdot \JJJ \right ].
\label{eqn7}
\end{equation}
It is easiest to work in an eigenbasis of $\KKK$ and (as will argue
later) we restrict the path integral to the row space of $\KKK$,
this gives
\begin{equation}
Z = \int_{-\infty}^{\infty} \ldots \int_{-\infty}^{\infty} d\QQ_1
\ldots d\QQ_{N-1}  \exp \left [\sum_{j=1}^{N-1} \left (-\frac 12
\QQ_j^2 a_j + \JJ_j \QQ_j \right ) \right ] \label{eqn8}
\end{equation}
where $\QQ_j$ are the coordinates associated with the eigenbasis of
$\KKK$ and $\QQ_N$ is associated with eigenvalue zero, $a_j$ is the
eigenvalue of $\KKK$  corresponding to $\QQ_j$, and $\JJ_j$ are the
components of $\JJJ$ in the eigenbasis of $\KKK$. The solution of
(\ref{eqn8}) is
\begin{equation}
Z = \left ( \frac {(2\pi)^{N-1}}{\prod_{j=1}^{N-1} a_j} \right )^{1/2} \prod_{j=1}^{N-1} \exp \left ( \frac {\JJ_j^2}{2a_j} \right ).
\label{eqn9}
\end{equation}
On our view, the experiment is described fundamentally by $\KKK$ and
$\JJJ$ on our topological graph. Again, per (\ref{eqn9}), there is
no field $\QQ$ appearing in $Z$ at this level, i.e., $\QQ$ is only
an integration variable.  $\QQ$ makes its first appearance as
something more than an integration variable when we produce
probabilities from $Z$. That is, since we are working with a
Euclidean path integral, $Z$ is a partition function and the
probability of measuring $\QQ_k=\QQ_0$ is found by computing the
fraction of $Z$ which contains $\QQ_0$ at the $k^{\mbox{th}}$
vertex\cite{lisi}. We have
\begin{equation}
P \left ( \QQ_k = \QQ_0 \right ) = \frac {Z \left ( \QQ_k = \QQ_0
\right )}{Z} = \sqrt{\frac {a_k}{2\pi}} \exp \left ( - \frac 12
\QQ_0^2 a_k + \JJ_k \QQ_0 - \frac {\JJ_k^2}{2a_k} \right )
\label{eqn10}
\end{equation}
as the part of theory X approximated in the continuum by QFT. The
most probable value of $\QQ_0$ at the $k^{\mbox{th}}$ vertex is then
given by
\begin{equation}
\delta P \left ( \QQ_k = \QQ_0 \right ) = 0 \Longrightarrow \delta
\left ( - \frac 12 \QQ_0^2 a_k + \JJ_k \QQ_0 - \frac {\JJ_k^2}{2a_k}
\right ) = 0 \Longrightarrow a_k \QQ_0 = \JJ_k. \label{eqn11}
\end{equation}
That is, $\KKK \cdot \QQQ_0 = \JJJ$ is the part of theory X that
obtains statistically and is approximated in the continuum by
classical field theory. We note that the manner by which $\KKK \cdot
\QQQ_0 = \JJJ$ follows from $P(\QQ_k = \QQ_0) = Z(\QQ_k = \QQ_0)/Z$
parallels the manner by which classical field theory follows from
QFT via the stationary phase method\cite{zee4}. Thus, one may obtain
classical field theory by the continuum limit of $\KKK \cdot \QQQ_0
= \JJJ$ in theory X (theory X $\rightarrow$ classical field theory),
or by first obtaining QFT via the continuum limit of $P(\QQ_k =
\QQ_0) = Z(\QQ_k = \QQ_0)/Z$ in theory X and then by using the
stationary phase method on QFT (theory X $\rightarrow$ QFT
$\rightarrow$ classical field theory). In either case, QFT is not
quantized classical field theory in our approach. In summary:

\begin{enumerate}
\item $Z$ is a partition function for an experiment described topologically by $\KKK/2+ \JJJ$  (Figure \ref{fig1}a).
\item $P(\QQ_k = \QQ_0) = Z(\QQ_k = \QQ_0)/Z$ gives us the probability for a particular geometric outcome in that experiment (Figures \ref{fig1}b and \ref{fig2}b).
\item $\KKK\cdot \QQQ_0 = \JJJ$ gives us the most probable values of the experimental outcomes which are then averaged to produce the geometry for the experimental procedure at the classical level (Figure \ref{fig2}a).
\item $P(\QQ_k = \QQ_0) = Z(\QQ_k = \QQ_0)/Z$ and $\KKK\cdot \QQQ_0 = \JJJ$ are the parts of theory X approximated in the continuum by QFT and classical field theory, respectively.
\end{enumerate}

\subsubsection{The Two-Source Euclidean Symmetry Amplitude/Partition
Function}

\begin{figure}
\includegraphics[height=50mm]{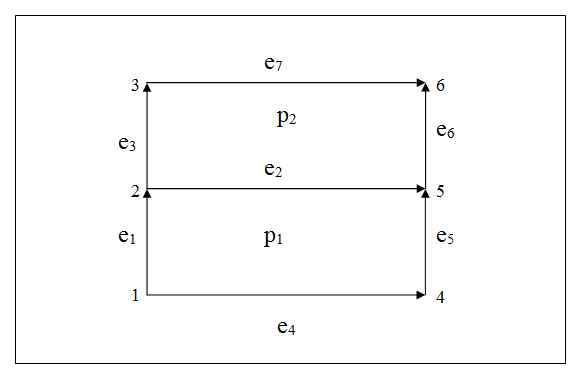}
\caption{Graph with six vertices, seven links $e_i$ and two
plaquettes $p_i$.} \label{fig3}
\end{figure}

Typically, one identifies fundamentally interesting physics with
symmetries of the action in the Central Identity of Quantum Field
Theory, but we have theory X fundamental to QFT, so our method of
choosing fundamentally interesting physics must reside in the
topological graph of theory X. Thus, we seek a constraint of $\KKK$
and $\JJJ$ in our graphical symmetry amplitude $Z$ and this will be
in the form of a self-consistency criterion (SCC). In order to
motivate our general method, we will first consider a simple graph
with six vertices, seven links and two plaquettes for our
$(1+1)-$dimensional spacetime model (Figure \ref{fig3}). Our goal
with this simple model is to seek relevant structure that might be
used to infer an SCC. We begin by constructing the boundary
operators over our graph.

The boundary of $\bfp_1$ is $\bfe_4 + \bfe_5 - \bfe_2 - \bfe_1$,
which also provides an orientation. The boundary of $\bfe_1$ is
$\bfv_2 - \bfv_1$, which likewise provides an orientation. Using
these conventions for the orientations of links and plaquettes we
have the following boundary operator for $C_2 \rightarrow C_1$,
i.e., space of plaquettes mapped to space of links in the spacetime
chain complex:
\begin{equation}
\partial_2 = \left [ \begin{array}{rr}
-1 & 0 \\
-1 & 1 \\
 0 & -1 \\
 1 & 0 \\
 1 & 0 \\
 0 & 1 \\
 0 & -1 \end{array} \right ]
\label{eqn12}
\end{equation}
Notice the first column is simply the links for the boundary of
$\bfp_1$ and the second column is simply the links for the boundary
of $\bfp_2$. We have the following boundary operator for $C_1
\rightarrow C_0$, i.e., space of links mapped to space of vertices
in the spacetime chain complex:
\begin{equation}
\partial_1 = \left [ \begin{array}{rrrrrrr}
-1 & 0 & 0 & -1 & 0 & 0 & 0 \\
 1 & -1 & -1 & 0 & 0 & 0 & 0 \\
 0 & 0 & 1 & 0 & 0 & 0 & -1 \\
 0 & 0 & 0 & 1 & -1 & 0 & 0 \\
 0 & 1 & 0 & 0 & 1 & -1 & 0 \\
 0 & 0 & 0 & 0 & 0 & 1 & 1 \end{array} \right ]
\label{eqn13}
\end{equation}
which completes the spacetime chain complex, $C_0 \leftarrow C_1
\leftarrow C_2$. Notice the columns are simply the vertices for the
boundaries of the edges. These boundary operators satisfy
$\partial_1\cdot\partial_2 = 0$, i.e., the boundary of a boundary
principle.

\begin{figure}
\includegraphics[height=30mm]{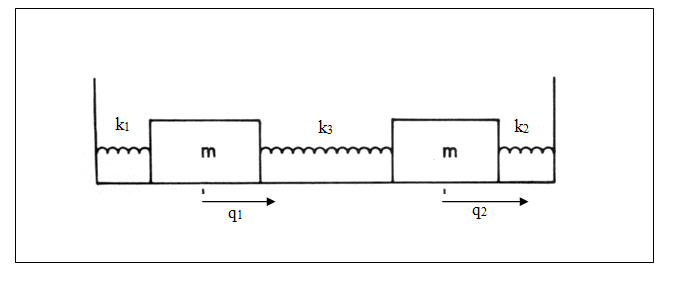}
\caption{Coupled harmonic oscillators.} \label{fig4}
\end{figure}

The potential for coupled oscillators can be written
\begin{equation}
V(q_1,q_2) = \sum_{a,b} \frac 12 k_{ab} q_a q_b = \frac 12 k q_1^2 + \frac 12 k q_2^2 + k_{12} q_1 q_2
\label{eqn14}
\end{equation}
where $k_{11} = k_{22} = k>0$ and $k_{12} = k_{21}<0$ per the
classical analogue (Figure \ref{fig4}) with $k = k_1 + k_3 = k_2 +
k_3$ and $k_{12} = -k_3$ to recover the form in (\ref{eqn14}).
The Lagrangian is then
\begin{equation}
L = \frac 12 m \dot{q}_1^2 + \frac 12 m \dot{q}_2^2 - \frac 12
kq_1^2 - \frac 12 k q_2^2 - k_{12} q_1q_2 \label{eqn15}
\end{equation}
so our NRQM Euclidean symmetry amplitude is
\begin{equation}
Z = \int Dq(t) \exp \left [ - \int_0^T dt \left ( \frac 12 m
\dot{q}_1^2 + \frac 12 m \dot{q}_2^2 + V(q_1, q_2) - J_1 q_1 - J_2
q_2 \right )\right ] \label{eqn16}
\end{equation}
after Wick rotation. This gives
\begin{equation}
\KKK = \left [ \begin{array}{rrrrrr}
\left ( \frac m{\Delta t} + k \Delta t \right ) & -\frac m{\Delta t} & 0 & k_{12} \Delta t & 0 & 0 \\
-\frac m{\Delta t} & \left ( \frac {2m}{\Delta t} + k \Delta t \right ) & -\frac m{\Delta t} & 0 & k_{12} \Delta t & 0 \\
0 & -\frac m{\Delta t} & \left ( \frac m{\Delta t} + k \Delta t \right ) & 0 & 0 & k_{12} \Delta t \\
k_{12} \Delta t & 0 & 0 & \left ( \frac m{\Delta t} + k \Delta t \right ) & -\frac m{\Delta t} & 0 \\
0 & k_{12} \Delta t & 0 & -\frac m{\Delta t} & \left ( \frac {2m}{\Delta t}  + k \Delta t \right ) & -\frac m{\Delta t} \\
0 & 0 & k_{12} \Delta t & 0 & -\frac m{\Delta t} & \left ( \frac m{\Delta t} + k \Delta t \right ) \end{array} \right ]
\label{eqn17}
\end{equation}
on our graph. Thus, we borrow (loosely) from Wise\cite{wise} and
suggest $\KKK \propto \partial_1\cdot\partial_1^T$ since
\begin{equation}
\partial_1\cdot\partial_1^T = \left [ \begin{array}{rrrrrr}
2 & -1 & 0 & -1 & 0 & 0 \\
-1 & 3 & -1 & 0 & -1 & 0 \\
0 & - 1& 2 & 0 & 0 & -1 \\
-1 & 0 & 0 & 2 & -1 & 0 \\
0 & -1 & 0 & -1 & 3 & -1 \\
0 & 0 & -1 & 0 & -1 & 2 \end{array} \right ]
\label{eqn18}
\end{equation}
produces precisely the same form as (\ref{eqn17}) and quantum theory
is known to be ``rooted in this harmonic paradigm''\cite{zee5}. [In
fact, these matrices will continue to have the same form as one
increases the number of vertices in Figure \ref{fig3}.] Now we
construct a suitable candidate for $\JJJ$, relate it to $\KKK$ and
infer our SCC.

Recall that $\JJJ$  has a component associated with each vertex so
here it has components, $J_n$, $n = 1, 2, \ldots, 6$; $J_n$ for $n =
1, 2, 3$ represents one source and $J_n$ for $n = 4, 5, 6$
represents the second source. We propose $\JJJ \propto
\partial_1\cdot\eee$, where $e_i$ are the links of our graph, since
\begin{equation}
\partial_1\cdot\eee =
\left [ \begin{array}{rrrrrrr}
-1 & 0 & 0 & -1 & 0 & 0 & 0 \\
 1 & -1 & -1 & 0 & 0 & 0 & 0 \\
 0 & 0 & 1 & 0 & 0 & 0 & -1 \\
 0 & 0 & 0 & 1 & -1 & 0 & 0 \\
 0 & 1 & 0 & 0 & 1 & -1 & 0 \\
 0 & 0 & 0 & 0 & 0 & 1 & 1 \end{array} \right ]
\left [ \begin{array}{c} e_1 \\ e_2 \\ e_3 \\ e_4 \\ e_5 \\ e_6 \\ e_7 \end{array} \right ]
 = \left [\begin{array}{c} -e_1-e_4 \\ e_1 - e_2-e_3 \\ e_3 - e_7 \\ e_4 - e_5 \\ e_2 + e_5 - e_6 \\ e_6 + e_7 \end{array}\right ]
\label{eqn19}
\end{equation}
automatically makes $\JJJ$ divergence-free, i.e., $\displaystyle
\sum_i J_i = 0$, and relationally defined. Such a relationship on
discrete spacetime lattices is not new. For example, Sorkin showed
that charge conservation follows from gauge invariance for the
electromagnetic field on a simplicial net\cite{sorkin2}.

With these definitions of $\KKK$  and $\JJJ$  we have, \emph{ipso
facto}, $\KKK\cdot \vvv \propto \JJJ$  as the basis of our SCC since
\begin{equation}
\partial_1\cdot\partial_1^T \cdot \vvv = \left [ \begin{array}{rrrrrr}
2 & -1 & 0 & -1 & 0 & 0 \\
-1 & 3 & -1 & 0 & -1 & 0 \\
0 & - 1& 2 & 0 & 0 & -1 \\
-1 & 0 & 0 & 2 & -1 & 0 \\
0 & -1 & 0 & -1 & 3 & -1 \\
0 & 0 & -1 & 0 & -1 & 2 \end{array} \right ] \left [
\begin{array}{c} v_1 \\ v_2 \\ v_3 \\ v_4 \\ v_5 \\ v_6 \end{array}
\right ] = \left [\begin{array}{c} -e_1-e_4 \\ e_1 - e_2-e_3 \\ e_3 - e_7 \\
e_4 - e_5\\ e_2 + e_5 - e_6 \\ e_6 + e_7 \end{array} \right ] =
\partial_1\cdot\eee \label{eqn20}
\end{equation}
where we have used $e_1 = v_2 - v_1$ (etc.) to obtain the last
column. You can see that the boundary of a boundary principle
underwrites (\ref{eqn20}) by the definition of ``boundary'' and from
the fact that the links are directed and connect one vertex to
another, i.e., they do not start or end `off the graph'. Likewise,
this fact and our definition of $\JJJ$ imply $\displaystyle \sum_i
J_i = 0$, which is our graphical equivalent of a divergence-free,
relationally defined source (every link leaving one vertex goes into
another vertex). Thus, the SCC $\KKK\cdot \vvv \propto \JJJ$ and
divergence-free sources $\displaystyle \sum_i J_i = 0$ obtain
tautologically via the boundary of a boundary principle. The SCC
also guarantees that $\JJJ$ resides in the row space of $\KKK$  so,
as will be shown, we can avoid having to ``throw away infinities''
associated with gauge groups of infinite volume as in Faddeev-Popov
gauge fixing. $\KKK$ has at least one eigenvector with zero
eigenvalue which is responsible for gauge invariance, so {\em the
self-consistent co-construction of space, time and divergence-free
sources entails gauge invariance.}

Moving now to $N$ dimensions, the Wick rotated version of (\ref{eqn3}) is (\ref{eqn6})
and the solution is (\ref{eqn7}). Using $\JJJ = \alpha
\partial_1\cdot\eee$  and $\KKK = \beta \partial_1\cdot
\partial_1^T$  ($\alpha, \beta \in \mathbb{R}$) with the SCC gives
$\KKK\cdot \vvv = (\beta/\alpha) \JJJ$, so that $\vvv =
(\beta/\alpha) \KKK^{-1}\cdot \JJJ$. However, $\KKK^{-1}$  does not
exist because $\KKK$ has a nontrivial null space, therefore the row
space of $\KKK$ is an $(N-1)-$dimensional subspace of the
$N-$dimensional vector space\footnote{This assumes the number of
degenerate eigenvalues always equals the dimensionality of the
subspace spanned by their eigenvectors.}. The eigenvector with
eigenvalue of zero, i.e., normal to this hyperplane, is $\left[
\begin{array}{ccccc} 1 & 1 & 1 & \ldots & 1 \end{array} \right ]^T$,
which follows from the SCC as shown supra. Since $\JJJ$ resides in
the row space of $\KKK$ and, on our view, $Z$ is a functional of
$\KKK$ and $\JJJ$ which produces a partition function for the
various $\KKK/2+\JJJ$  associated with different 4D experimental
configurations, we restrict the path integral of (\ref{eqn6}) to the
row space of $\KKK$. Thus, our approach revises (\ref{eqn7}) to give
(\ref{eqn9}).

Since this is linear, we do not expect to recover GR in this manner.
Instead, we expect to make correspondence with GR via a modification
to Regge calculus, a form of lattice gravity.

\section{Recovering General Relativity: RBW Versus Hiley's Implicate Order}
\label{section3}

The modeling of ``undivided wholeness'' (monism) in each formalism
leads to the same problem for both approaches when dealing with GR,
i.e., how to relate/connect different M4 frames. This is simply to
say the essence of gravity in GR is spacetime curvature, i.e., the
relative acceleration of `neighboring' geodesics, whereas the other
forces are modeled via deviation from geodetic motion in a flat
spacetime. Consider, for example, the phenomenon of gravitational
lensing that produces an Einstein ring image of a distant quasar by
an intervening galaxy. The explanation per GR is that empty
spacetime around the worldtube of the intervening galaxy is curved
so that null geodesics near its worldtube are deformed or `bent'
thereby `lensing' the photons as they proceed from the quasar around
the galaxy to Earth. We note that the principle explanatory
mechanism, i.e., spacetime curvature, doesn't have anything to do
with the stress-energy tensor of the quasar, or of the photons
passing through that region of space, or of Earth. Yet, the monistic
view doesn't allow for a separation of this sort - if we're relating
the quasar, galaxy, photons, and Earth, then the stress-energy
tensor for all these objects must be produced together with the
geometry of spacetime from a single `entity'.

For Hiley, this `entity' will be a process-based algebra of the
implicate order. Specifically, he speculates, a deformed Poisson
algebra obtained via the hidden Heisenberg algebra gives the Moyal
algebra for relating phase information for our electromagnetic
interactions. If he proceeds with a current algebra approach (again,
as inferred by his approach to Schr\"{o}dinger, Pauli and Dirac
particles), presumably, he will have to promote the spacetime metric
to a field so that it will have its own particle and current. Then,
he will have to produce commutation relations between the
electromagnetic current and the gravitational current to describe
the possible outcomes at interaction vertices. The problem is, of
course, there are no spacetime locations for the interaction
vertices, since one result of the calculation itself must be the
spacetime geometry. Of course, if this algebra produces dual
currents as with the Dirac particle, one is again left with the
problem of figuring out which currents correspond to actual detector
outcomes. But, suppose he takes the hint from his Dirac result and
gives up on the idea of ``Bohmian trajectories,'' as he has with the
``Bohmian guidance equation,''
\cite{callaghan}\cite{hiley14}\cite{hiley15} and proceeds with a
canonical quantization. Since his shadow manifolds are particular
Lorentz frames rather than the full M4 for the Dirac equation, the
logical counterpart to his approach (if it exists) for GR would be a
particular foliation of the curved spacetime manifold. That is, a
shadow manifold would be a particular path through all possible
three geometries and matter fields in the solution space of H = 0.

For RBW, the single `entity' responsible for its monism is
spacetimematter and we note immediately that for us the GR
explanation of the Einstein ring in the above example must be
corrected since there is no ``empty spacetime.'' Thus, per RBW, GR
is only an approximation to the `correct' theory of gravity. Of
course this is not new, the same can be said of Newtonian gravity
given GR and Newtonian mechanics given special relativity. The
questions are, what is the `correct' theory of gravity and in what
sense is it approximated by GR? Since our underlying approach is
graphical, we start with the graphical version of GR, called Regge
calculus, and propose modifications thereto.

\begin{figure}
\includegraphics[height=70mm]{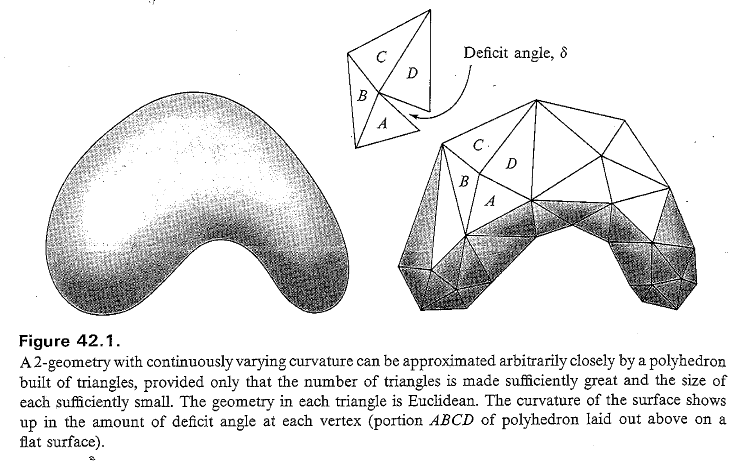}
\caption{Reproduced from Misner, C.W., Thorne, K.S., Wheeler, J.A.:
Gravitation. W.H. Freeman, San Francisco (1973), p. 1168. Permission
pending.} \label{fig5}
\end{figure}

In Regge calculus, the spacetime manifold is replaced by a lattice
geometry where each cell is Minkowskian (flat). Typically, this
lattice spacetime is viewed as an approximation to the continuous
spacetime manifold, but the opposite could be true and that is what
we will advocate. The lattice reproduces a curved manifold as the
cells (typically 4D `tetrahedra' called ``simplices'') become
smaller (Figure \ref{fig5}). Curvature is represented by ``deficit
angles'' (Figure \ref{fig5}) about any plane orthogonal to a
``hinge'' (triangular side to a tetrahedron, which is a side of a
simplex). A hinge is two dimensions less than the lattice dimension,
so in 2D a hinge is a zero-dimensional point (Figure \ref{fig5}).
The Hilbert action for a vacuum lattice is $I_R = \frac{1}{8
\pi}\displaystyle \sum_{\sigma_i\in L}\varepsilon_{i} A_{i}$ where
$\sigma_i$ is a triangular hinge in the lattice \emph{L}, $A_i$ is
the area of $\sigma_i$ and $\varepsilon_i$ is the deficit angle
associated with $\sigma_i$. The counterpart to Einstein's equations
is then obtained by demanding $\frac {\delta
I_R}{\delta\ell_{j}^{2}}=0$ where $\ell_{j}^{2}$ is the squared
length of the $j^{th}$ lattice edge, i.e., the metric. To obtain
equations in the presence of matter-energy, one simply adds the
matter-energy action $I_{M}$ to $I_R$ and carries out the variation
as before to obtain $\frac {\delta I_R}{\delta\ell_{j}^{2}}= -\frac
{\delta I_{M}}{\delta\ell_{j}^{2}}$. One finds the stress-energy
tensor is associated with lattice edges, just as the metric, and
Regge's equations are to be satisfied for any particular choice of
the two tensors on the lattice. Thus, Regge's equations are, like
Einstein's equations, a self-consistency criterion for the
stress-energy tensor and metric.

It seems to us that the most glaring deviation from GR phenomena
posed by directly connected sources per theory X would be found in
the exchange of photons on cosmological scales. Therefore, using
Regge calculus, we constructed a Regge differential equation for the
time evolution of the scale factor $a(t)$ in the Einstein-de Sitter
cosmology model (EdS) and proposed two modifications to the Regge
calculus approach: (1) we allowed the graphical links on spatial
hypersurfaces to be large, as when the interacting sources reside in
different galaxies, and (2) we assumed luminosity distance $D_L$ is
related to graphical proper distance $D_p$ by the equation $D_L =
(1+z)\sqrt{\overrightarrow{D_p}\cdot \overrightarrow{D_p}}$, where
the inner product can differ from its usual trivial
form\cite{stuckey3}. There are two reasons we made this second
assumption. First, in our view, space, time and sources are
co-constructed, yet $D_p$ is found without taking into account EM
sources responsible for $D_L$. That is to say, in Regge EdS (as in
EdS) we assume that pressureless dust dominates the stress-energy
tensor and is exclusively responsible for the graphical notion of
spatial distance $D_p$. However, even though the EM contribution to
the stress-energy tensor is negligible, EM sources are being used to
measure the spatial distance $D_L$. Second, in our view, there are
no ``photon paths being stretched by expanding space,'' so we cannot
simply assume $D_L = (1+z)D_p$ as in EdS. The specific form of
$\KKK\cdot \QQQ_0 = \JJJ$ that we used to find the inner product for
$D_L$ was borrowed from linearized gravity in the harmonic gauge,
i.e., $\partial^2 h_{\alpha\beta} = -16 \pi G (T_{\alpha\beta} -
\frac{1}{2} \eta_{\alpha\beta} T)$. That is, $D_L = (1+z)\sqrt{1 +
h_{11}}D_p$ and we use $\KKK\cdot \QQQ_0 = \JJJ$ to find $h_{11}$.
We emphasize that $h_{\alpha\beta}$ here corrects the graphical
inner product $\eta_{\alpha\beta}$ in the inter-nodal region between
the worldlines of photon emitter and receiver, where
$\eta_{\alpha\beta}$ is obtained via a matter-only stress-energy
tensor. Since the EM sources are negligible in the matter-dominated
solution and we're only considering a classical deviation from a
classical background, we have $\partial^2 h_{\alpha\beta} = 0$ to be
solved for $h_{11}$. Obviously, $h_{11} = 0$ is the solution that
gives the trivial relationship, but allowing $h_{11}$ to be a
function of $D_p$ allows for the possibility that $D_L$ and $D_p$
are not trivially related. We have $h_{11} = AD_p + B$ where $A$ and
$B$ are constants and, if the inner product is to reduce to
$\eta_{\alpha\beta}$ for small $D_p$, we have $B = 0$. Presumably,
$A$ should follow from the corresponding theory of quantum gravity,
so an experimental determination of its value provides a guide to
quantum gravity per our view of classical gravity. As we will show,
our best fit to the Union2 Compilation data gives $A^{-1}$ = 8.38
Gcy, so the correction to $\eta_{11}$ is negligible except at
cosmological distances, as expected.

\begin{figure}[h]
\begin{center}
\includegraphics[height=60mm]{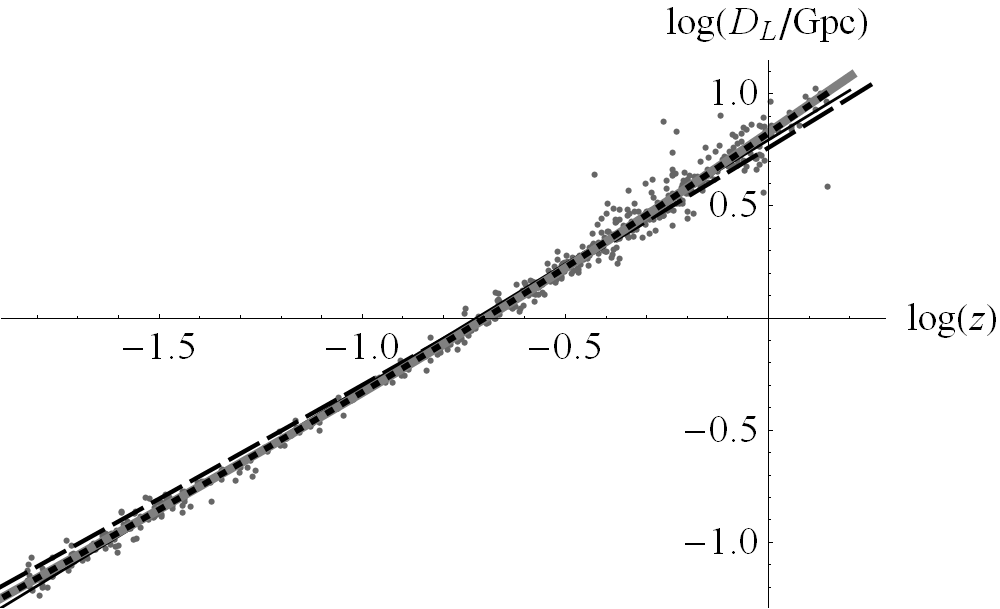}
\end{center}
\caption{Plot of transformed Union2 data along with the best fits for linear
regression (thin black), EdS (dashed), $\Lambda$CDM (gray), and MORC (dotted).} \label{fig8}
\end{figure}

\begin{figure}[h]
\begin{center}
\includegraphics[height=60mm]{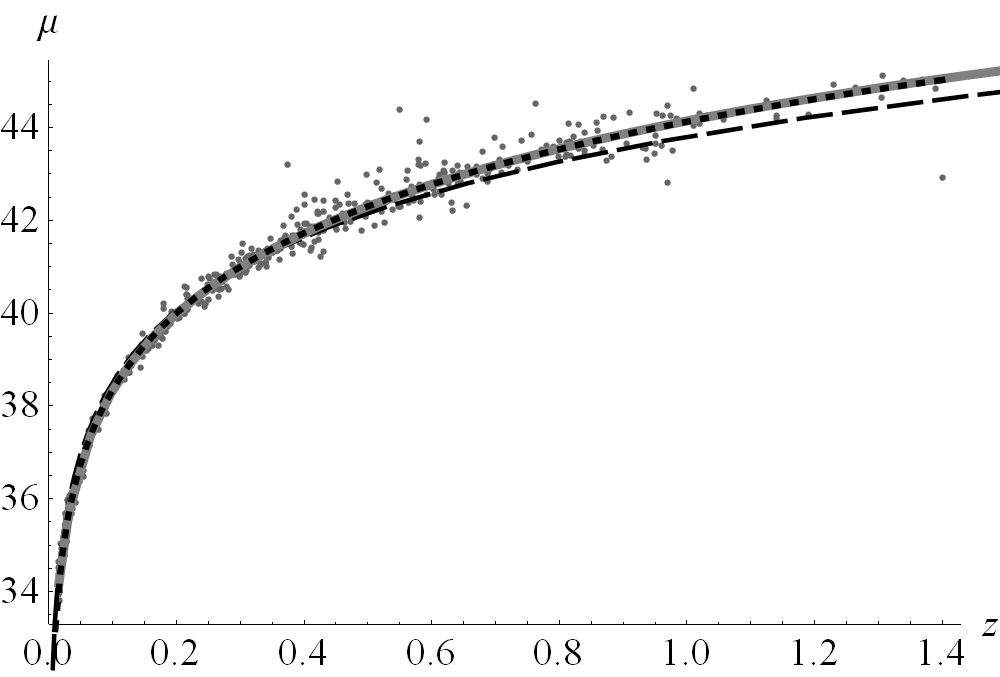}
\end{center}
\caption{Plot of Union2 data along with the best fits for EdS (dashed),
$\Lambda$CDM (gray), and MORC (dotted).  The MORC curve is terminated at $z$ = 1.4 in this figure
so that the $\Lambda$CDM curve is visible underneath.}
\label{fig9}
\end{figure}

The modified Regge calculus model (MORC), EdS and the concordance
model $\Lambda$CDM (EdS plus a cosmological constant to account for
dark energy) were compared using the data from the Union2
Compilation, i.e., distance moduli and redshifts for type Ia
supernovae\cite{amanullah} (see Figures \ref{fig8} and \ref{fig9}).
We found that a best fit line through $\displaystyle
\log{\left(\frac{D_L}{\mbox{Gpc}}\right)}$ versus $\log{z}$ gives a
correlation of 0.9955 and a sum of squares error (SSE) of 1.95. By
comparison, the best fit $\Lambda$CDM gives SSE = 1.79 using $H_o$ =
69.2 km/s/Mpc, $\Omega_{M}$ = 0.29 and $\Omega_{\Lambda}$ = 0.71.
The parameters for $\Lambda$CDM yielding the most robust fit to
``the Wilkinson Microwave Anisotropy Probe data with the latest
distance measurements from the Baryon Acoustic Oscillations in the
distribution of galaxies and the Hubble constant
measurement\cite{komatsu}'' are $H_o$ = 70.3 km/s/Mpc, $\Omega_{M}$
= 0.27 and $\Omega_{\Lambda}$ = 0.73, which are consistent with the
parameters we find for its Union2 Compilation fit. The best fit EdS
gives SSE = 2.68 using $H_o$ = 60.9 km/s/Mpc. The best fit MORC
gives SSE = 1.77 and $H_o$ = 73.9 km/s/Mpc using $R = A^{-1}$ = 8.38
Gcy and $m = 1.71\times 10^{52}$ kg, where $R$ is the coordinate
distance between nodes, $A^{-1}$ is the scaling factor from our
non-trival inner product explained above, and $m$ is the mass
associated with nodes\footnote{Strictly speaking, the stress-energy
tensor is associated with graphical links, not nodes. Our
association of mass with nodes is merely conceptual.}. A current
(2011) ``best estimate'' for the Hubble constant is $H_o$ = (73.8
$\pm$ 2.4) km/s/Mpc \cite{riess2}. Thus, MORC improves EdS as much
as $\Lambda$CDM in accounting for distance moduli and redshifts for
type Ia supernovae even though the MORC universe contains no dark
energy is therefore always decelerating.

This is but one test of the RBW approach and MORC must pass more
stringent tests in the context of the Schwarzschild solution where
GR is well confirmed. However, MORC's empirical success in dealing
with dark energy gives us reason to believe this formal approach to
classical gravity may provide creative new techniques for solving
other long-standing problems, e.g., quantum gravity, unification,
and dark matter. In particular, if MORC passes empirical muster in
the context of the Schwarzschild solution, then information such as
$A^{-1}$ might provide guidance to a theory of quantum gravity
underlying a graphical classical theory of gravity.

\section{The Problems of Time: RBW versus the Implicate
Order on Being and Becoming} \label{section4}
\subsection{The Implicate Order}

It is obvious that a process conception of fundamental reality does
not sit well with blockworld or frozen time. In the case of
blockworld there is no unique `now' successively coming into
existence. There are an indenumerably infinite number of time-like
foliations of M4, each representing a unique global `now' at various
values of its foliating time, and a particular spatial hypersurface
in foliation A (a `now' for observer A) contains events on many
different spatial hypersurfaces in foliation B (different `nows' for
observer B). That events which are simultaneous for observer A are
not simultaneous for observer B is called the ``relativity of
simultaneity'' and negates an objective passage of time. That is to
say, there is no objective (frame independent) distinction in
spacetime between past, present and future events respectively and
therefore no objective distinction to be had about the occurrence or
non-occurrence of events. In the words of Costa de
Beauregard\cite{beauregard}:

\begin{quotation}This is why first
Minkowski, then Einstein, Weyl, Fantappi\`{e}, Feynman, and many
others have imagined space-time and its material contents as spread
out in four dimensions. For those authors, of whom I am one ...
relativity is a theory in which everything is ``written'' and where
change is only relative to the perceptual mode of living
beings.\end{quotation}

And we have seen that the canonical or gauge interpretation of GR
leads to an even ``blockier'' world than SR! As Earman puts
it\cite{earman}:

\begin{quotation}Taken at face value, the gauge interpretation of
GTR implies a truly frozen universe: not just the `block universe'
that philosophers endlessly carp about -- that is, a universe
stripped of A-series change or shifting `nowness' -- but a universe
stripped of its B-Series change in that no genuine physical
magnitude (= gauge invariant quantity) changes its value with
time.\end{quotation}

As for the problem of frozen time in canonical QG, as we said, the
dynamics of the theory are given by a Hamiltonian operator
$\hat{H}$, which is defined on a space of spin network states via
the equation $\hat{H}|\Psi\rangle=0$, i.e., the Wheeler-DeWitt
equation mentioned earlier. It is hard to see how to avoid the
problem of frozen time in canonical QG because, unlike the standard
Schrödinger equation
$\hat{H}|\Psi\rangle=i\hbar\frac{\partial|\Psi\rangle}{\partial t}$,
the RHS of the Wheeler-DeWitt equation disappears. Because time is
part of the physical system being quantized, there is no external
time with respect to which the dynamics could unfold, only the
analogous gauge symmetries are there.

Therefore, in order to preserve his process model of reality, at the
end of the day Hiley must end up with a fundamental physical theory
that avoids the blockworld of relativity and the frozen time of
canonical QG. We can only speculate as to exactly how Hiley will
address these concerns or even exactly how his program will recover
GR, therefore the reader should consider our suggestions tentative.
Let's discuss SR first and extrapolate from there.

Any argument from SR to blockworld requires, as a premise, realism
about the geometric properties of M4. As we indicated earlier, we
think Hiley might be in a position to reject such realism because in
his scheme each shadow manifold constitutes a particular Lorentz
frame. Every Lorentz observer will construct his own space and time.
These space-times can exist together, but we cannot ascribe a
sharply defined `time' as to when they all exist together. Therefore
every frame is its own coordinate origin of its own explicate
manifold. Think of the implicate order as the head of an octopus and
the various explicate shadow manifolds (e.g., individual
perspectives or proper times) as the many tentacles produced from
the implicate order by the holomovement. Every event is described by
an infinity of times and spatial locations even though there is only
one event that all Lorentz observers are observing, as related
formally by the Lorentz group. The shadow manifolds are not
connected directly and thus there is no M4 as conceived by
Minkowski-there isn't one spacetime. As for the problem of time in
canonical GR and in canonical QG, again, assuming he gives up on
Bohmian trajectories and guide waves and uses canonical quantization
per his yet-to-be-determined process algebra for gravity and all
other forces, then his shadow manifolds correspond to particular
paths through all possible three geometries and matter fields in the
solution space of H = 0. Thus, Hiley avoids ``frozen time'' in GR
and QG exactly like he avoids it in SR -- by giving up on the idea
of a unique explicate order \emph{a la} M4, leaving the unification
of perspective to the implicate order as dictated by the
holomovement. Whether or not such a view is Hiley's considered view
and whether or not it is any better off than solipsism, we do not
know. Hiley is clear however that blockworld defined as the reality
of all events past, present and future is inconsistent with his
process ontology. This means he either rejects realism about M4 at
its root or provides a physically and formally acceptable preferred
foliation in addition to the structure of M4.

Given that Hiley rejects blockworld it would be reasonable to assume
that he embraces some form of presentism (only the present is real).
However in his theory of moments\cite{hiley17} he clearly rejects
presentism. According to Hiley's theory of moments, the holomovement
gives rise to ``moments/durons'' (which involves information from
the past and the future). Of moments he goes on to say that: ``For a
process with a given energy cannot be described as unfolding at an
instant except in some approximation''\cite{hiley18}. As we
understand it, the idea is that the holomovement can explicate
either a small region or a large region of spacetime (to include the
future) `at once', though never the entire universe. The extent of
the explicate domain (how much of the future exists) depends on the
properties of the holomovement in each particular case and the
process is apparently stochastic. In Hiley's model therefore, just
as the past can effect what unfolds in the future, so the future can
influence what unfolds in the present and what unfolded in the past.
Hiley is clear that what happens in the future cannot be made to
rewrite the past, but that the future possibilities can influence
the unfolding of the present. What is less clear is whether these
moments pass in and out of existence or always stay in existence
once explicated. All this suggests that each individual shadow
manifold is constantly changing in its own time (evolving `now')
such that the past is consistent with the present and the future is
understood probabilistically. Again, the solipsistic view of
individual shadow manifolds connected via the implicate order per
the holomovement avoids the blockworld implication of M4. One could
imagine other hybrid models of blockworld and presentism (or at
least becoming) such as entire blockworld universes winking
discretely in and out existence, each one different in some way from
the last. How to formalize models such as these, whether in an
algebraic program or some other, is unclear to us. What is clear to
us at the end of the day, merely advocating for fundamental physics
based on process isn't enough to secure every feature of dynamism.
Whether or not quantum theory and relativity can be unified in such
a way as to uphold all of dynamism is a formal question that has yet
to be resolved.

\subsection{RBW}
Of course we happily accept the implication of relativity theory
that it is a block universe and we are not bothered by the problem
of frozen time in canonical QG because we reject dynamism at its
foundation. For those wedded to dynamism these results are puzzling
embarrassments that require some sort of compatibilist response or a
completely new process-based ontology and formalism. In RBW we start
at bottom with an adynamical global constraint, a self-consistency
criterion (SCC) that allows us to construct discrete spacetimematter
graphs from which all the other effective theories and their
concomitant phenomena emerge. According to RBW, what quantum theory
and relativity theory are both trying to tell us is that every facet
of dynamism is false. If we succeed in our program of unification,
we will have shown that nothing in physics itself demands dynamism,
rather it was just a historical contingency based in the fact that
all physics must start with experience. Perhaps RBW offers a fourth
possibility regarding the nature of time, i.e., time as part of a
fundamental (pregeometric) regime wherein the notions of space, time
and matter are co-defined and co-determining. Technically, time,
space and matter as stand-alone concepts are not fundamental,
emergent or illusions in RBW. We note that it is only from a God's
eye Point of view (the view from nowhere and nowhen) that time and
change are an illusion and in a fundamentally relational model such
as ours there are no perspectives ``external to the universe.'' The
conceptual foundation of our dynamical reality isn't a so-called
``initial singularity,'' but the adynamical SCC upon which all
dynamic theories reside. The SCC characterizing spacetimematter at
the bottom of RBW is not a dynamical law or initial condition, but
it is responsible for the discrete action. Therefore, if
higher-level physical theories are truly recovered from the discrete
action, then there is nothing left to explain at bottom, regardless
what phenomena one counts as initial/boundary conditions versus
laws. The point of all this is that in RBW there will be no quantum
cosmology as is currently conceived. We also note that the universe
comes with many physically significant modes of temporal passage and
change such as proper time, cosmic time, etc. Certainly these
constitute objective notions of becoming (objectively dynamical
flow) even if they are mere patterns in a block universe. Therefore,
RBW does not negate change and becoming, it merely internalizes and
relativizes them.

Of course, all this falls short of getting every facet of time as
experienced into fundamental physics. There is no objectively
distinguished present moment, and there is no objectively dynamical
becoming in the sense of bringing events into existence that never
existed before from a God's eye point of view. However, perhaps the
standard wisdom that time as experienced is either a physical
feature of reality or merely a psychological feature of conscious
beings is a false dichotomy. Perhaps what all this suggests is that
conscious temporal experience is fundamental as well, so instead of
spacetimematter at bottom we have the super-monistic
spacetimematterexperience at bottom. This is shear speculation of
course, it would require working out a new formal model and much
else conceptually. We can say however that the alternatives are not
very appetizing if we take the frozen block universe seriously. The
image of consciousness crawling along the worldtube of individuals
illuminating the present and moving it toward the future is an
unhelpful and non-explanatory kind of dualism which simply exempts
conscious experience from the rules of the block
universe\cite{weyl}. The other alternative, that conscious
experience emerges from or is realized in neuro-dynamical activity,
is problematic in a block universe in which everything, past,
present and future is just there `at once' (including conscious
experiences throughout the block) and brains are just worldtubes
like everything else. One might find correlations between brain
states and the experience of the objective specialness of the `now'
and the experience of objectively dynamical becoming, but it cannot
be said that brain dynamics produce or bring into being conscious
states (themselves worldtubes). In such a universe brain processes
are not metaphysically or causally more fundamental than conscious
processes. Again, the idea of spacetimematterexperience is
half-baked, but if we take it seriously, perhaps it moves RBW closer
to Hiley's process conception of reality since process (objectively
distinguished present and objectively dynamical) is the nature of
ordinary conscious experience and the experience of time partially
motivates the process model.

\bibliographystyle{spphys}

\end{document}